\documentclass[
  journal=medium,
  manuscript=,
  year=2023,
  volume=Preprint,
]{cup-journal}

\usepackage{amsmath}
\usepackage[nopatch]{microtype}
\usepackage{booktabs}

\title{Human Protein–Protein Interaction Networks: A Topological Comparison Review}

\author{Rodrigo Henrique Ramos}
\affiliation{University of São Paulo, São Carlos, SP, Brazil}
\alsoaffiliation{Federal Institute of São Paulo, São Carlos, SP, Brazil}
\email[Rodrigo Henrique Ramos]{ramos@ifsp.edu.br | rodrigohenrique.ramos@usp.br}

\author{Cynthia de Oliveira Lage Ferreira}
\affiliation{University of São Paulo, São Carlos, SP, Brazil}

\author{Adenilso Simao}
\affiliation{University of São Paulo, São Carlos, SP, Brazil}

\keywords{protein–protein interaction networks, PPIN, complex systems, centrality measures, network topology} 

\begin{document}

\begin{abstract}

Protein-Protein Interaction Networks aim to model the interactome, providing a powerful tool for understanding the complex relationships governing cellular processes. These networks have numerous applications, including functional enrichment, discovering cancer driver genes, identifying drug targets, and more. Various databases make protein-protein networks available for many species, including \textit{Homo sapiens}. This work topologically compares four \textit{Homo sapiens} networks using a coarse-to-fine approach, comparing global characteristics, sub-network topology, specific nodes centrality, and interaction significance. Results show that the four human protein networks share many common protein-encoding genes and some global measures, but significantly differ in the interactions and neighbourhood. Small sub-networks from cancer pathways performed better than the whole networks, indicating an improved topological consistency in functional pathways. The centrality analysis shows that the same genes play different roles in different networks. We discuss how studies and analyses that rely on protein-protein networks for humans should consider their similarities and distinctions.


\end{abstract}

\section{Introduction}
Complex networks are characterized by intricate structures and involve multiple nodes interacting in potentially unknown ways. These networks are subject to internal and external factors that can lead to self-organization and emergent phenomena~\autocite{barabasi2002}.

The creation of Protein-Protein Interactions Networks (PPINs) became possible thanks to the advancements in large-scale methods to identify the functional relationship between genes. These methods include gene expression correlations, protein-protein interactions, text mining associations, and others~\autocite{Network_analysis_2016_EMBL}. Protein interactions play a crucial role in numerous biological processes, and analyzing their network can reveal unexpected biology, shedding light on complex pathways and potential therapeutic targets~\autocite{Network_analysis_2012_Tutorial, ProteinsInteractions2014}. With the mapping of most humans' genes, genomics has made progress in identifying genetic variations and their correlation with diseases. However, understanding how these variations interfere with complex biological functions with thousands of interactions remains a significant challenge~\autocite{ProteinsInteractions2014}.

Researchers have utilized the analysis of PPINs and their functional modules to study complex diseases such as cancer and diabetes~\autocite{PPIN_and_disease_2014}. This has resulted in the development of computational methods aimed at improving our understanding of these diseases, including identifying driver mutations in cancer~\autocite{jorge_driver_review}.

Motivated by the broad applicability and use of PPINs and the existence of several databases that provide PPINs that aim to model the human interactome, this study explores the similarities and topological differences between PPINs following a coarse-to-fine approach. We analyzed global metrics, cancer sub-networks, interaction significance, intersections between nodes and edges, and the centrality of nodes. The results reveal that PPINs have similar global features, particularly in cancer sub-networks, but differ in measures related to the neighborhood and mainly in the intersection of edges.

This work is organized as follows. The next section describes the databases and the pre-processing made considering the interaction significance. Afterwards, we analyze the nodes and edge intersection. The fourth and fifth section compares the PPINs and cancer sub-networks using global measures. After that, we explore the centrality role of three sets of nodes in the four PPINs. Finally, we present the results and the concluding remarks.



\section{Databases and Pre-Processing}

The combination of scientific literature curation and computational methods has led to the creation of numerous protein interaction databases~\autocite{Protein_protein_interaction_databases_2015}. Although there were doubts about the quality of interaction data at the beginning, significant advances in approaches over the past decade have greatly improved the data quality. Additionally, the standardization of data and curation practices across databases has enhanced the confidence and availability of interaction data~\autocite{ProteinsInteractions2014}. The International Molecular Exchange consortium\footnote{http://www.imexconsortium.org/} (IMEx) strives to standardize curation rules related to interaction identification, while also establishing a standard data format, site search interface, and free access through the Creative Commons Attribution License.

In this study, we have chosen four human PPINs databases:  HINT~\autocite{HINT}, IntAct~\autocite{IntAct}, Reactome~\autocite{Reactome}, and STRING~\autocite{STRING}. These databases were chosen based on their frequent updates and citation rates. While IntAct and STRING have PPINs for multiple organisms, we have limited the networks to only \textit{Homo sapiens}. Different databases may have varying methods of linking two proteins, but they assign a confidence score (i.e., edge weight) to each connection, which ranges from zero to one. This score represents the likelihood that the association is accurate~\autocite{STRING}. HINT (High-quality INTeractomes) is an exception to this rule since its interactions come from a consensus between databases, keeping only significant interactions. Thus, its edges have no weight.

\begin{figure}[!htb]
    \vspace*{-3mm}
    \centering
    \includegraphics[width=0.9\linewidth]{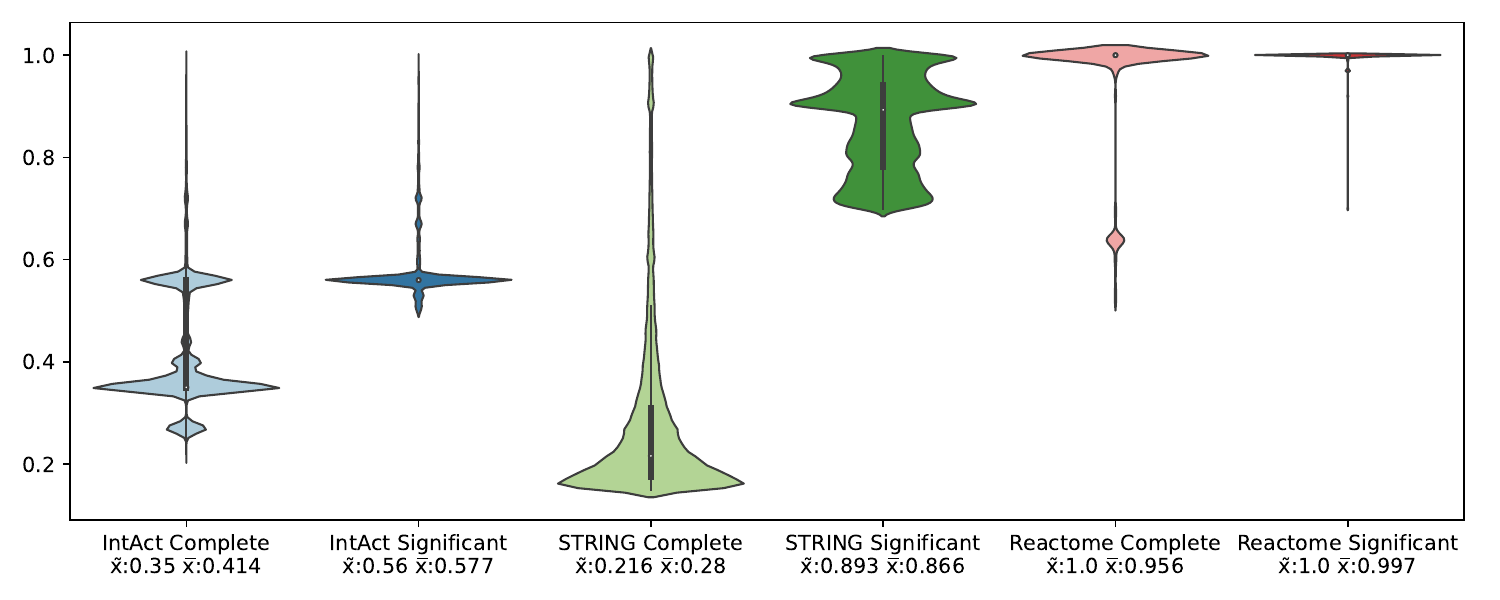}
    \caption{Edge score distribution for all networks with scores. The median $\tilde{x}$ and mean $\overline{x}$ are closer in significant networks, normalizing the distribution. STRING shows the biggest difference between the significant and complete versions, since most of the edges score in the complete network are below 40\%.}
    \label{fig:Edge Score Distribution}
\end{figure}

We followed HINT's method to enhance the quality of IntAct, Reactome, and STRING by creating significant versions. This was done by eliminating edges with low scores. A comparison of the edge score distribution before and after the removal is shown in Figure~\ref{fig:Edge Score Distribution}. The complete versions of IntAct and STRING have the majority of their interactions with a confidence score below 50\%, with STRING having a median of 28\%. Reactome, on the other hand, has very few low-significant interactions, with a median of 96\%. To create significant networks for Reactome and STRING, we removed interactions with scores below 70\%. For IntAct, since there are barely any interactions with scores above 60\%, we removed edges with a confidence score of 50\% or less.

Figure~\ref{fig:Edge Score Distribution} shows how different the PPINs are when considering the significance of their interactions. Table~\ref{tab:networksNodesAndEdges} shows another important difference between the PPINs: the number of nodes and edges. STRING Complete has almost six million interactions; however, as shown in Figure~\ref{tab:networksNodesAndEdges}, most of these interactions have very low confidence scores. IntAct Complete has the second highest number interactions, but is 18 times lower than STRING Complete. This shows how important it is to remove low-confidence interactions, especially when comparing the PPINs.


\begin{table}[hbt!]
\begin{threeparttable}
\caption{Number of Nodes and Edges in Complete and Significant Networks. In the complete networks, the nodes range from 12,184 to 20,807, and the edges from 257,629 to 5,968,680.}
\label{tab:networksNodesAndEdges}
\begin{tabular}{lll}
\toprule
\headrow Network & Nodes  & Edges \\ \midrule
HINT                 & 15,386 & 119,494   \\ \midrule
IntAct Complete      & 20,807 & 328,288   \\ \midrule
IntAct Significant   & 13,437 & 91,701     \\ \midrule
Reactome Complete    & 13,953 & 257,629   \\ \midrule
Reactome Significant & 12,184 & 228,699   \\ \midrule
STRING Complete      & 19,382 & 5,968,680 \\ \midrule
STRING Significant   & 16,812 & 252,953   \\ \bottomrule
\end{tabular}
\end{threeparttable}
\end{table}

After removing low-confidence scores, we eliminate self-loop edges and keep only the largest connected component. This pre-processing step is necessary because some network analysis measures require a single connected component and no self-loops. Reactome has no self-loop, and STRING has only one. HINT has 4\% of its edges as self-loops, while IntAct has 1\%. The largest connected component in each network has similar sizes, on average keeping 97.75\% of its nodes. Table~\ref{tab:finalNetworks} shows the number of nodes and edges after the pre-processing steps. 

\begin{table}[hbt!]
\begin{threeparttable}
\caption{Final Networks. The networks present a more evenly distribution among nodes and edges while maintaining only significative interactions.}
\label{tab:finalNetworks}
\begin{tabular}{lll}
\toprule
\headrow Network & Nodes  & Edges \\ \midrule
HINT     & 14,763 & 114,588   \\ \midrule
IntAct   & 13,268 & 90,446     \\ \midrule
Reactome & 11,873 & 228,447   \\ \midrule
STRING   & 16,582 & 252,801  \\ \bottomrule
\end{tabular}
\end{threeparttable}
\end{table}

All the analyses made from now on consider the networks present in Table~\ref{tab:finalNetworks}. The final networks used in this work narrow the great difference in the number of nodes and edges among the original networks, making possible a fair comparison while keeping only interaction more likely to be true.


\section{Nodes and Edges Intersections}

PPINs are designed to represent the interactions within a cell. Even though these networks are regularly updated, the information they contain is still not comprehensive~\autocite{PPIN_and_disease_2014}. Various methods, such as link prediction, can be used to determine how two proteins interact with each other~\autocite{linkPredictionPPIN_2020}. This section delves into the intersections that exist in the nodes and edges of the four PPINs.

\begin{figure}[!htb]
    \centering
    \includegraphics[width=0.75\linewidth]{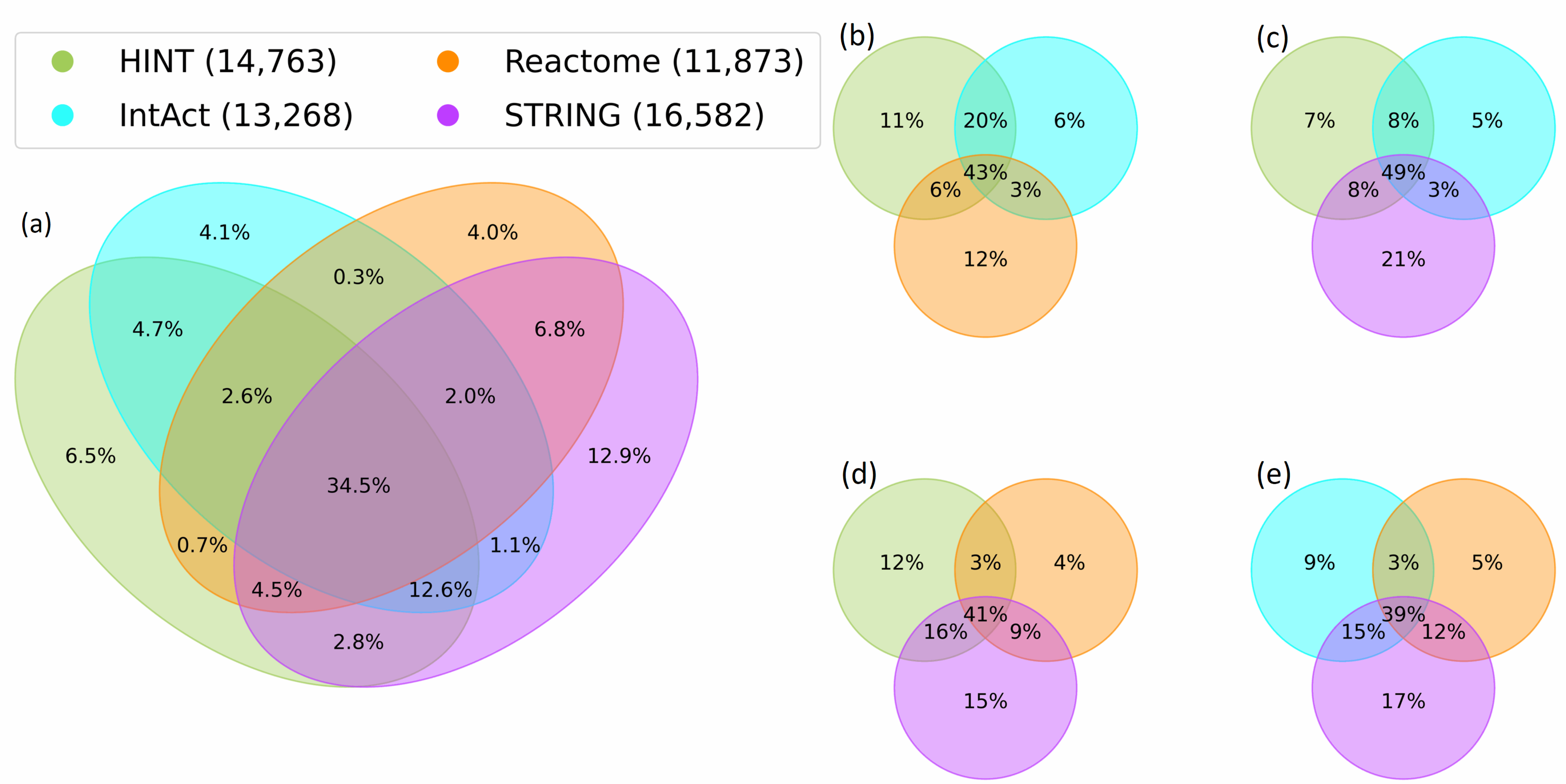}
    \caption{Nodes Intersection. All networks nodes intersection (a). Trio of networks nodes intersections (b), (c), (d), and (e). The networks have overlapping protein names while maintaining a significant amount of exclusive proteins.}
    \label{fig:Nodes Intersection}
\end{figure}

Figure~\ref{fig:Nodes Intersection} shows the nodes' intersection between the four networks and the intersection in groups of three. The union of all networks' nodes, Figure~\ref{fig:Nodes Intersection}-(a), corresponds to 21,475 unique nodes. From this union, 34,5\% (7,408 nodes) are present in the four networks. Non-intersecting areas in Figure~\ref{fig:Nodes Intersection}-(a) show the percentage of unique nodes in each network. STRING harbours 2,770 unique nodes (12.9\% ), while 1,460 (6.8\% ) of the nodes are present only in STRING and Reactome. Figure~\ref{fig:Nodes Intersection}-(b) reveals that without STRING, HINT and IntAct share 20\% of their nodes, the highest value when comparing groups of three. 

The disparity in the networks' number of nodes impacts the percentual intersection analysis. To address this issue, Table~\ref{tab:nodesIntersection} provides the percentage of nodes from each network (lines) included in the other networks (columns). HINT has 78\% of its nodes included in IntAct and 79\% in STRING. Even though the percentage is similar, STRING has more nodes than IntAct, indicating a greater intersection between HINT and IntAct than with HINT and STRING. The average column shows that around 3/4 of the nodes in each network are contained in other networks. Overall, Table~\ref{tab:nodesIntersection} shows that PPINs are still incomplete and in development. While there is some overlap between the networks, it still needs to be completed.

\begin{table}[hbt!]
\begin{threeparttable}
\caption{Network Nodes Contained in Other Networks.}
\label{tab:nodesIntersection}
\begin{tabular}{l|c|c|c|c|c}
\toprule
\text
 		& 		\textbf{HINT} & \textbf{IntAct} & \textbf{Reactome} & \textbf{STRING} & \textbf{Average}	\\ \bottomrule
\textbf{HINT}     &  	 & 78\% & 61\% &  79\%  & 73\%\\ \midrule
\textbf{IntAct}   & 88\% &  	& 64\% &  81\%  & 77\%\\ \midrule
\textbf{Reactome} & 76\% & 71\% &      &  86\%  & 78\%\\ \midrule
\textbf{STRING}   & 70\% & 65\% & 62\% &    	& 65\%\\ \midrule
\end{tabular}
\end{threeparttable}
\end{table}

\begin{figure}[!htb]
    \centering
    \includegraphics[width=.75\linewidth]{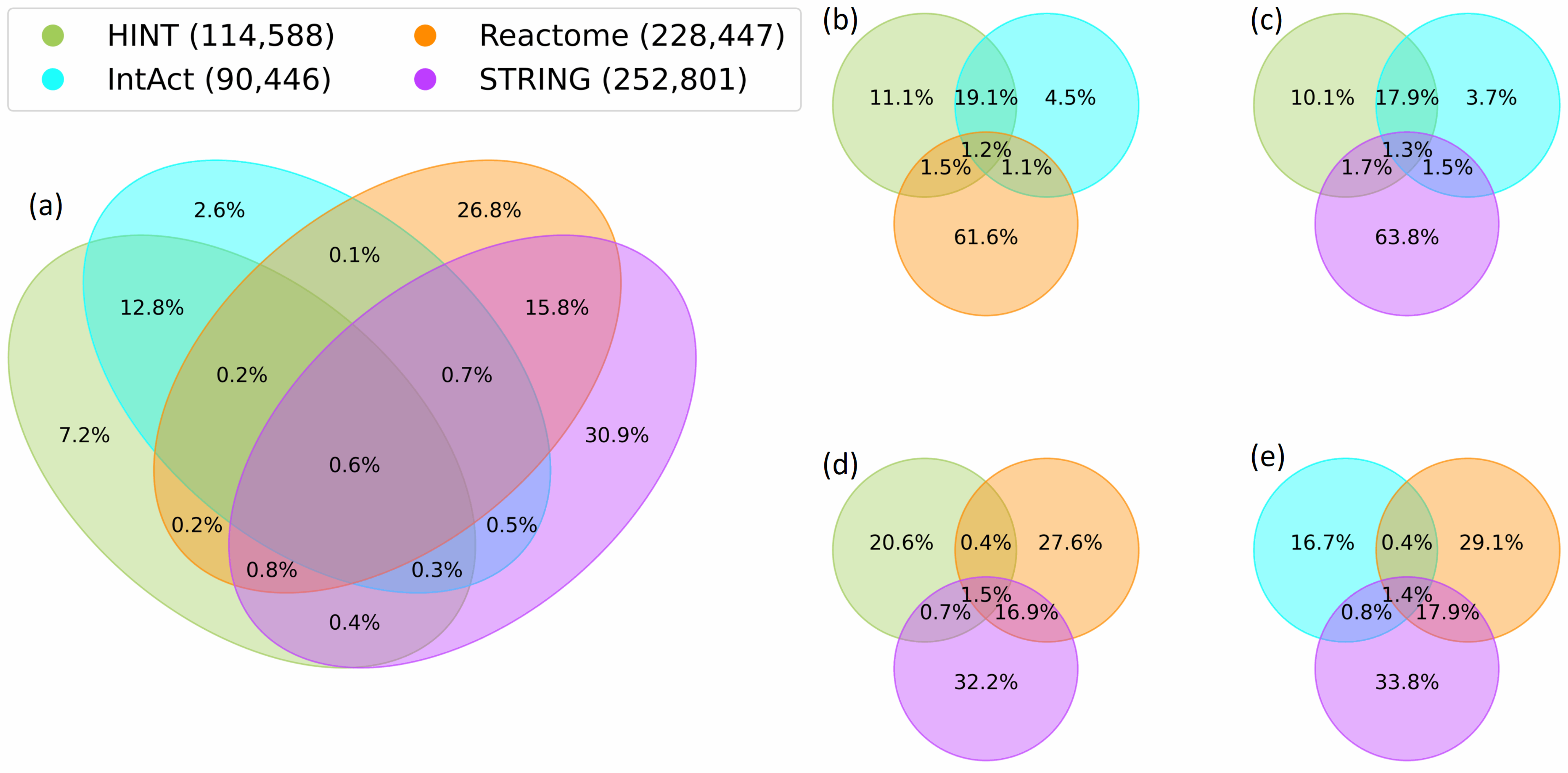}
    \caption{Edges Intersection. All networks edges intersection (a). Trio of networks edges intersections (b), (c), (d), and (e). The networks show minimal edges intersection.}
    \label{fig:Edges Intersection}
\end{figure}

We replicate the analysis made with nodes using edges. The results show significant differences. Figure~\ref{fig:Edges Intersection}-(a) reveals that only 0.6\% of the edges are shared among all networks. The in trio analyses also show a mutual overlap ranging from 1.2\% to 1.5\%. The number of edges in the networks has a greater difference than the number of nodes, which can affect the percentage analysis. Table~\ref{tab:edgesIntersection} shows a clear intersection pattern between HINT and IntAct, as well as Reactome and STRING.

\begin{table}[hbt!]
\begin{threeparttable}
\caption{Network Edges Contained in Other Networks.}
\label{tab:edgesIntersection}
\begin{tabular}{l|c|c|c|c|c}
\toprule
\text
 		& 		\textbf{HINT} & \textbf{IntAct} & \textbf{Reactome} & \textbf{STRING} & \textbf{Average}	\\ \bottomrule
\textbf{HINT}     &  	 & 62\% & 8\% &  10\%  & 27\% \\ \midrule
\textbf{IntAct}   & 78\% &  	& 9\% &  12\%  & 33\% \\ \midrule
\textbf{Reactome} & 4\%  & 4\%  &     &  40\%  & 16\% \\ \midrule
\textbf{STRING}   & 4\%  & 4\%  & 36\%&    	   & 15\% \\ \midrule
\end{tabular}
\end{threeparttable}
\end{table}

The previous results show a similarity between the nodes present in the networks but a significant variation in overlapping edges. We selected 10 popular genes~\autocite{10popularGenes} to analyze the edges between them and further compare the interactions overlap. There are 45 possible edge combinations. We present in Table~\ref{tab:Edges Between Popular Genes} only the 21 edges that appear in at least one of the networks. The total number of edges impacts the chance that an edge exists in a network, but STRING has 11\% more edges than Reactome while having twice as many "matches". The 10 selected nodes are considered hubs, as shown in Section~\ref{sec:centralities}, and the probability of edges between hubs is related to assortativity, a subject we will address in Figure~\ref{fig:assortatividade}.

\begin{table}[hbt!]
\begin{threeparttable}
\caption{Edges Between 10 Popular Genes. The table presents only pairs of genes interacting in at least one network.}
\label{tab:Edges Between Popular Genes}
\begin{tabular}{|l|c|c|c|c|c|c|c|c|c|c|c|}
\toprule
\headrow \textbf{Gene Pair}&
  \begin{tabular}[c]{@{}c@{}}APOE\\ ESR1\end{tabular} &
  \begin{tabular}[c]{@{}c@{}}APOE\\ IL6\end{tabular} &
  \begin{tabular}[c]{@{}c@{}}EGFR\\ AKT1\end{tabular} &
  \begin{tabular}[c]{@{}c@{}}EGFR\\ ESR1\end{tabular} &
  \begin{tabular}[c]{@{}c@{}}EGFR\\ IL6\end{tabular} &
  \begin{tabular}[c]{@{}c@{}}EGFR\\ VEGFA\end{tabular} &
  \begin{tabular}[c]{@{}c@{}}ESR1\\ AKT1\end{tabular} &
  \begin{tabular}[c]{@{}c@{}}IL6\\ AKT1\end{tabular} &
  \begin{tabular}[c]{@{}c@{}}TNF\\ AKT1\end{tabular} &
  \begin{tabular}[c]{@{}c@{}}TNF\\ APOE\end{tabular} &
  \begin{tabular}[c]{@{}c@{}}TNF\\ EGFR\end{tabular} \\ \bottomrule
\textbf{HINT}     &   &   & X &   &   &   &   &   &   &   &   \\ \midrule
\textbf{IntAct}   &   &   & X & X &   &   &   &   &   &   &   \\ \midrule
\textbf{Reactome} & X &   & X & X & X & X & X &   &   &   &   \\ \midrule
\textbf{STRING}   &   & X & X & X & X & X & X & X & X & X & X \\ \bottomrule
\headrow \textbf{Gene Pair} &
  \begin{tabular}[c]{@{}c@{}}TNF\\ IL6\end{tabular} &
  \begin{tabular}[c]{@{}c@{}}TNF\\ VEGFA\end{tabular} &
  \begin{tabular}[c]{@{}c@{}}TP53\\ AKT1\end{tabular} &
  \begin{tabular}[c]{@{}c@{}}TP53\\ EGFR\end{tabular} &
  \begin{tabular}[c]{@{}c@{}}TP53\\ ESR1\end{tabular} &
  \begin{tabular}[c]{@{}c@{}}TP53\\ IL6\end{tabular} &
  \begin{tabular}[c]{@{}c@{}}TP53\\ TNF\end{tabular} &
  \begin{tabular}[c]{@{}c@{}}TP53\\ VEGFA\end{tabular} &
  \begin{tabular}[c]{@{}c@{}}VEGFA\\ AKT1\end{tabular} &
  \begin{tabular}[c]{@{}c@{}}VEGFA\\ IL6\end{tabular} &
   \\ \bottomrule
\textbf{HINT}     &   &   & X &   &   &   &   &   &   &   &   \\ \midrule
\textbf{IntAct}   &   &   &   &   &   &   &   &   &   &   &   \\ \midrule
\textbf{Reactome} &   &   & X & X &   &   &   & X & X &   &   \\ \midrule
\textbf{STRING}   & X & X & X & X & X & X & X & X & X & X &   \\ \bottomrule
\end{tabular}
\end{threeparttable}
\end{table}

Table~\ref{tab:Edges Between Popular Genes} reinforces the finding that, albeit the networks share a considerable number of nodes, the interaction between them varies greatly. Just the edge EGDR-AKT1 is present in all networks, while 11 edges are only in STRING.

\section{Global Networks Measures}
In this section, we present six global measures to explore and compare the networks as a whole. Firstly, we analyze the degree distribution and characterize the networks as scale-free using the Power Law Package~\autocite{powerLawPackage2014}. 

\begin{figure}[!htb]
    \centering
    \includegraphics[width=1\linewidth]{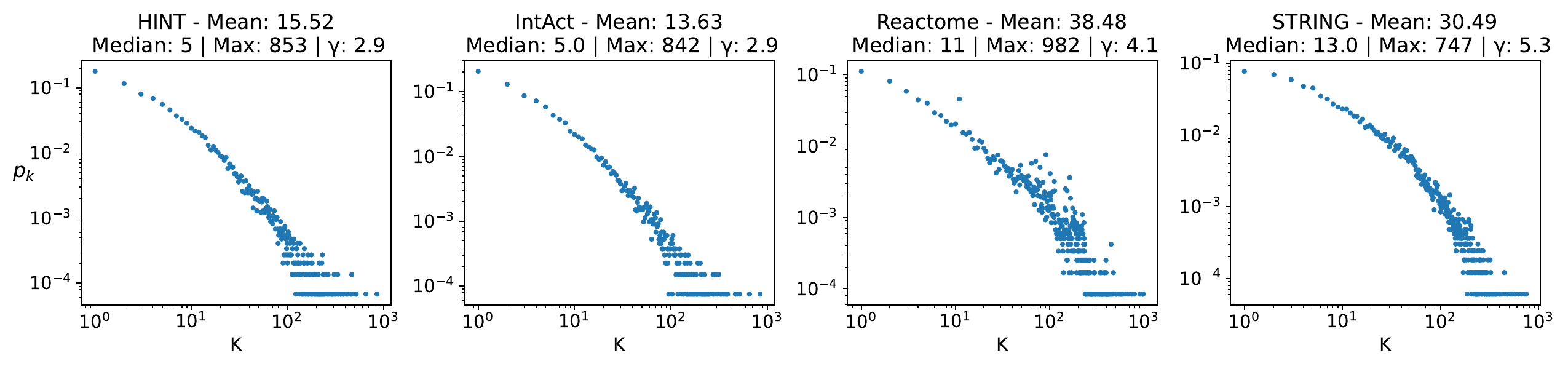}
    \caption{Scale Free Characterization. The four PPINs share a similar scale free degree distribution, especially the pairs HINT \& IntAct and Reactome \& STRING.}
    \label{fig:scaleFree}
\end{figure}

A key difference between random and real-world networks is their degree distribution. Real-world networks tend to have a scale-free distribution where few nodes are highly connected (hubs), while most nodes have few neighbours~\autocite{barabasi2015networkScience}. Thus, the mean and variance cannot capture the distribution behaviour. While not all real-world networks are scale-free, the ones that are follow a degree distribution probability similar to a power law. This means that, in scale-free networks, it is more likely to find small-degree nodes than high-degree nodes. Figure \ref{fig:scaleFree} shows the degree distribution probability, in log-log scale, for the PPINs. After the network name, there is information about the degree distribution.

The average degree is more than double the median, showing that most nodes have a small degree and few hubs are responsible for increasing the average value. The value Max shows the degree of the biggest hub. Even though STRING has the most edges, it has the smaller Max value. The last information, $\gamma$ shows the approximate exponent of a power-law distribution that best represents the degree distribution of each network~\autocite{powerLawPackage2014}. Although the PPINs have varying numbers of nodes and edges and differ in their intersections, they share similarities in their degree distribution. The plots demonstrate a similar trend, with Reactome appearing slightly less defined in the middle. The plots' scales are also nearly identical. These observations classify all four PPINs as scale-free networks, where the majority of nodes have a small degree, and only a few hubs have a much higher degree than the average. 

\begin{figure}[!htb]
    \centering
    \includegraphics[width=1\linewidth]{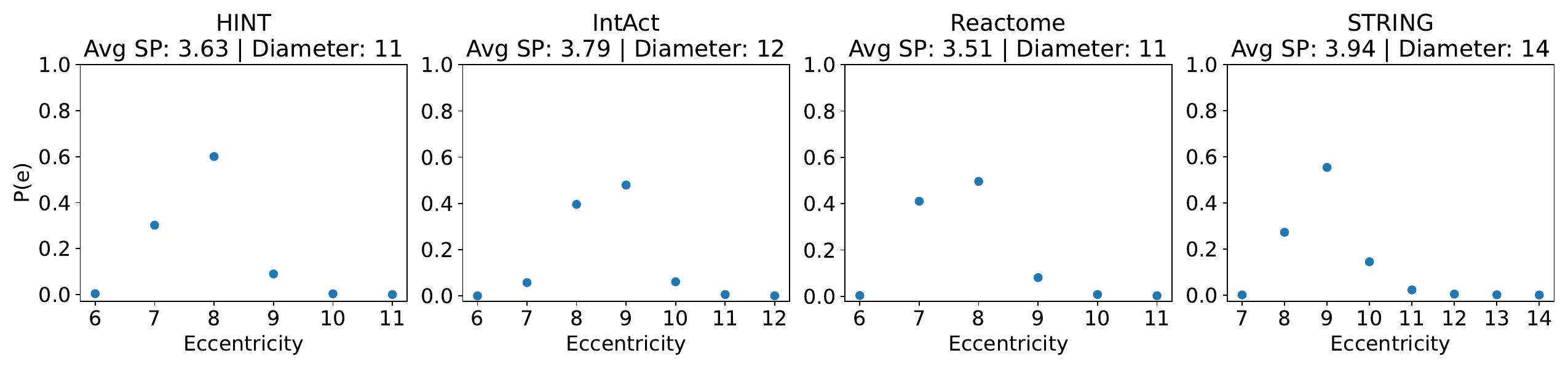}
    \caption{Small World. The four PPINs show a small world behaviour, even with thousands of nodes, the average shortest path is smaller than 4.}
    \label{fig:smallWorld}
\end{figure}

One of the consequences expected from scale-free networks is also being considered a small world. The term ``small-world'' is used to describe networks that, despite having thousands of nodes, have a relatively small diameter and eccentricity distribution. The eccentricity of a node refers to the maximum shortest path from that node to all other nodes, while the diameter is the maximum eccentricity found in the network. Figure~\ref{fig:smallWorld} displays the eccentricity probability distribution, diameter and average shortest path of the PPINs.

The behaviour of the four networks regarding the small world phenomena is quite similar overall. The average shortest path (Avg SP) ranges from 3.51 to 3.94, indicating that, on average, information takes less than four ``steps'' to move through the network. Except for STRING, the diameter is also similar, ranging from 11 to 12. STRING has 16,582 nodes, with the shortest path between its two farthest nodes being 14. Looking at the probability distribution for eccentricity, we can see that in STRING, almost 60\% of nodes have an eccentricity of 9, and less than 10\% have an eccentricity greater than 12, indicating that the network's diameter is a scenario for a few nodes.  

\begin{figure}[!htb]
    \centering
    \includegraphics[width=1\linewidth]{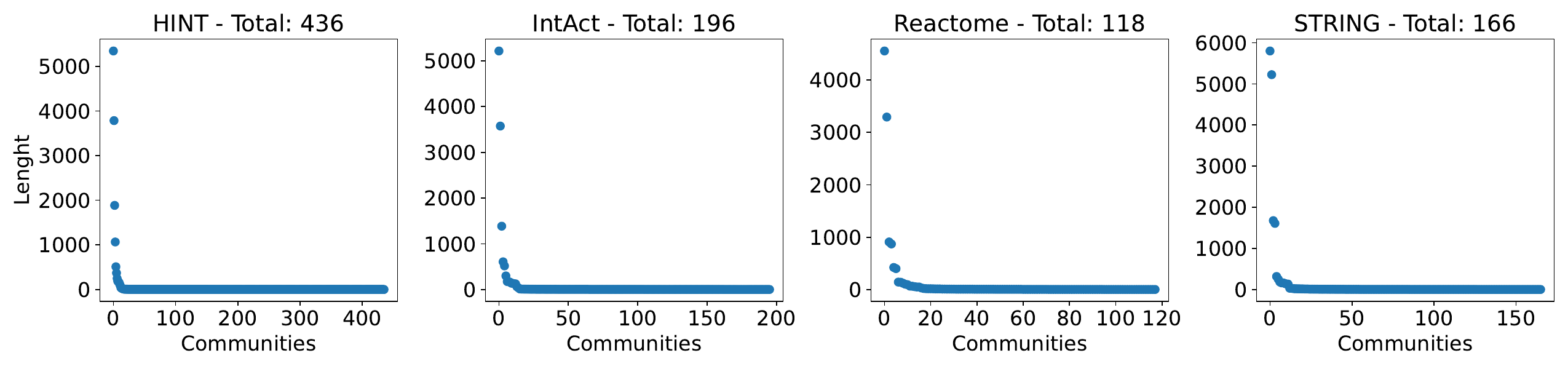}
    \caption{Communities. The community length distribution size is similar, with few large and many small communities.}
    \label{fig:communities}
\end{figure}

Groups of proteins work together to accomplish complex tasks~\autocite{ProteinsInteractions2014}. These groups form clusters and communities, with larger clusters often associated with functional modules, such as pathways \autocite{Network_analysis_2016_EMBL}. The study of PPIN and its functional modules has been utilized to investigate complex diseases like cancer and diabetes \autocite{Network_analysis_2012_Tutorial,PPIN_and_disease_2014}. Figure~\ref{fig:communities} shows the total number of communities found in the PPINs, along with their length distribution.

\begin{figure}[!htb]
    \centering
    \includegraphics[width=1\linewidth]{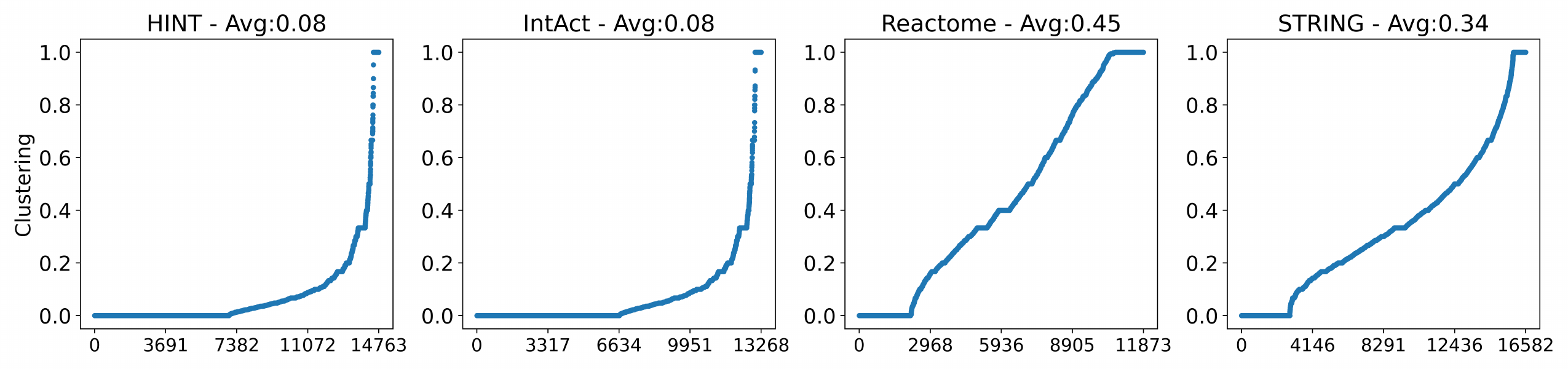}
    \caption{Clustering. The pair HINT \& IntAct have almost equal distribution. Reactome \& STRING are similar, while being distinct from HINT \& IntAct.}
    \label{fig:clustering}
\end{figure}

Just as it happened with the small world analysis, the PPINs show an overall similar behaviour in communities. The total number of communities in IntAct, Reactome, and STRING ranges from 118 to 196. All networks have very few large modules, with the vast majority having less than 100 nodes. HINT, for example, has 5,344 nodes in its largest community, while 425 communities have less than 100 nodes. Since the communities are made from interconnections within the nodes, we also present the clustering distribution for the PPINs in Figure~\ref{fig:clustering}.

HINT and IntAct share the same average clustering and almost identical clustering distribution, with few nodes having 100\% clustering and nearly half having zero clustering. Reactome has the most interconnections between nodes. While the PPINs share similarities in terms of communities, there are noticeable differences in node clustering.

\begin{figure}[!htb]
    \centering
    \includegraphics[width=1\linewidth]{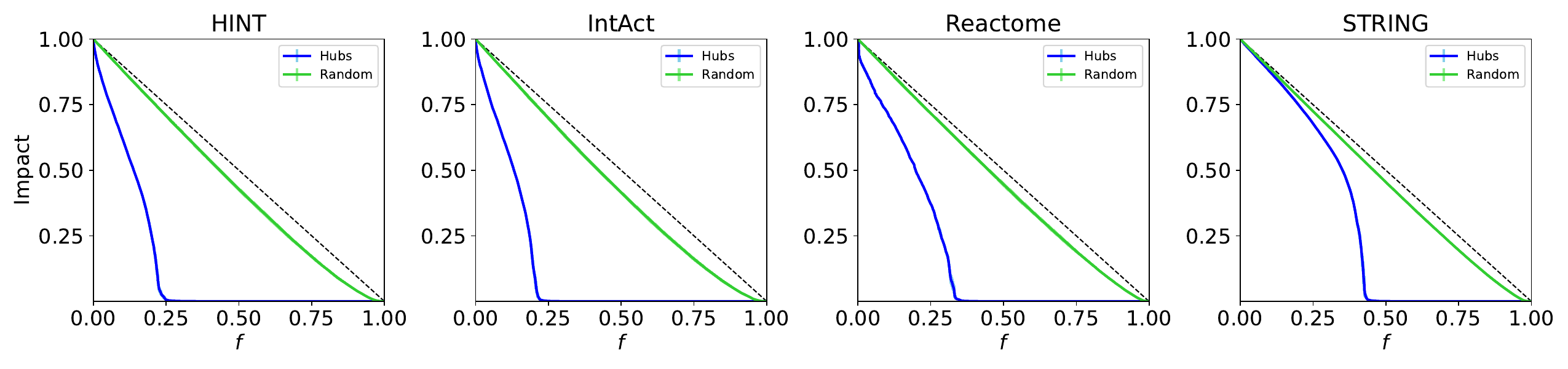}
    \caption{Network Resilience. The PPINs show a behaviour expected from scale free networks: resilient to random removal and fragile to hub removal. STRING is the most resilient due to its degree assortativity.}
    \label{fig:attack}
\end{figure}

In Figure~\ref{fig:attack}, the resilience of the network is examined through the removal of nodes via random and hub attacks. Random attacks test for failures that occur unintentional, whereas hub removal simulates planned attacks on important nodes.

On the graph, the x-axis ($f$) shows the fraction of nodes removed, while the y-axis shows the impact on the largest connected component. We conducted 30 random removal executions and 10 hub removal executions. A vertical line positioned over the impact line represents the standard deviation between these executions, indicating the variance in the results. In all cases, the variance is minimal. The dotted black line indicates zero impact, where removal does not break the network in more than one connected component. The four PPINs are very resilient to random removal and weak to hubs removal, as expected from scale free-networks~\autocite{barabasiAttack}. Out of the four, HINT and IntAct are the most vulnerable to hub attacks, as they become completely dysfunctional after 25\% of their hubs are removed. Reactome is more robust than HINT and IntAct, as it only reaches zero on the y-axis after 37\% of its hubs are removed. On the other hand, STRING shows remarkable resilience to hub attacks, with only around 45\% of its hubs removal leading to a zero y-axis value. The density and clustering of the networks play a critical role in their resilience, but compared to Reactome, STRING has lower values in both metrics. We attribute STRING's high resilience to its degree assortativity, which enables hubs to connect to other hubs and create many redundant paths within the network's core.

\begin{figure}[!htb]
    \centering
    \includegraphics[width=1\linewidth]{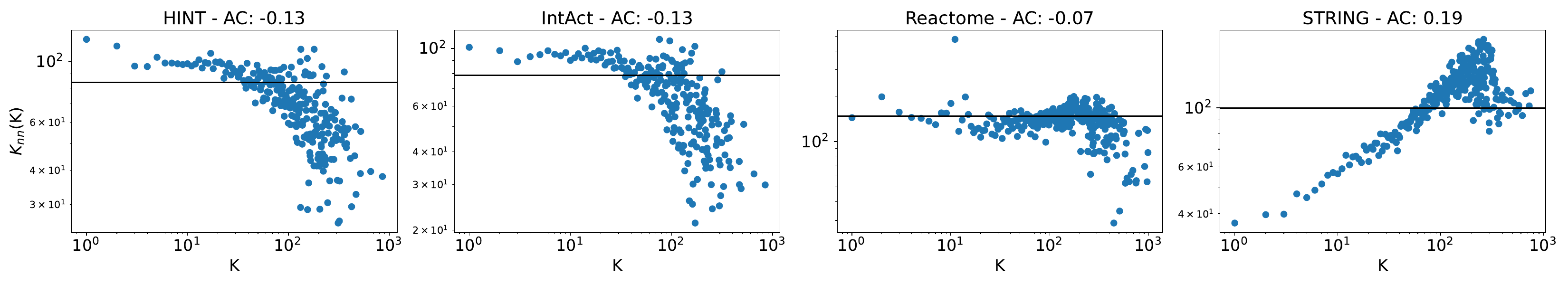}
    \caption{Assortativity. The PPINs do not follow a linear distribution. HINT \& IntAct follows the same trend, while Reactome and STRING have unique comportment.}
    \label{fig:assortatividade}
\end{figure}

In our last global network measure, we explored degree assortativity. This measure evaluates how nodes connect with other nodes based on their degree. Figure~\ref{fig:assortatividade} shows the assortativity coefficient (AC) after the network name. Theses values ranges from -0.13 to 0.19. Positive AC indicates that low degree nodes tends to connect with low degree nodes, and high degree nodes to connect with other high degree nodes. Negative AC indicates the opposite, where low degree nodes tend to connect with high degree. 

The AC alone does not capture the actual assortativity behaviour found in the PPINs. Thus, Figure~\ref{fig:assortatividade} also present the assortativity distribution. In the plot, $K$ (x-Axis) represents all nodes with degree $K$, and $K_{nn}(K)$ (y-Axis) is the average neighbours' degree of nodes with degree $K$. Each sub-plot's solid black horizontal line represents neutral assortativity (AC equals zero). HINT and IntAct have the same AC and a similar distribution. In both cases, nodes with degree smaller than forty show a neutral assortativity. As the degree increases, the plot chances tendency, indicating that hubs are connected with small degree nodes. Reactome keeps a neutral assortativity until around degree three hundred. Only the significant hubs show negative assortativity. Overall, the three PPINs share similarities: neutral assortativity within small degree nodes and negative assortativity on hubs. STRING has the opposite behaviour. Small degree nodes are connected with small degree nodes, showing a positive assortativity trend until degree two hundred and fifty. After that, big hubs tend to connect with other similar or smaller hubs.

\section{Cancer sub-networks}
This section compares the PPINs by analyzing sub-networks extracted from cancer pathways. Pathways consist of interacting genes that contribute to specific biological functions. These pathways serve as the foundational components of a cell's complex system~\autocite{Reactome2020}. The KEGG database~\autocite{KEGG} makes available cancer pathways, i.e., lists of genes associated with specific types of cancer. Using pathways for colorectal, pancreatic, and glioma cancer, we create induced subgraphs from the PPINs, totalizing twelve cancer pathway networks. Table~\ref{tab:pathways networks} presents these networks and five topological measures in order to explore how the same set of genes creates distinct sub-networks in different PPINs.

\begin{table}[!htb]
\caption{Topological Analysis for Cancer Pathways Networks}
\label{tab:pathways networks}
\begin{tabular}{|l|c|c|c|c|c|c|c|c|}
\hline
\rowcolor[HTML]{FFFFFF} 
                    & LCC (\%)  & AvgSP & AvgC & $\rho$ & Com\\ \hline
\rowcolor[HTML]{FFFFFF} 
Colorectal HINT     & 84\% & 3.32  & 0.20 & 0.06  & 6  \\ \hline
\rowcolor[HTML]{FFFFFF} 
Colorectal IntAct   & 76\% & 4.05  & 0.21 & 0.05  & 8  \\ \hline
\rowcolor[HTML]{FFFFFF} 
Colorectal Reactome & 100\% & 2.14  & 0.61 & 0.19 & 4  \\ \hline
\rowcolor[HTML]{FFFFFF} 
Colorectal STRING   & 100\% & 1.97  & 0.62 & 0.22 & 4  \\ \hline
\rowcolor[HTML]{C0C0C0} 
Pancreatic HINT     & 80\% & 3.01  & 0.26 & 0.08  & 6  \\ \hline
\rowcolor[HTML]{C0C0C0} 
Pancreatic IntAct   & 68\% & 4.05  & 0.16 & 0.06  & 7  \\ \hline
\rowcolor[HTML]{C0C0C0} 
Pancreatic Reactome & 100\% & 1.97  & 0.57 & 0.22 & 4  \\ \hline
\rowcolor[HTML]{C0C0C0} 
Pancreatic STRING   & 100\% & 1.96  & 0.61 & 0.22 & 4  \\ \hline
\rowcolor[HTML]{FFFFFF} 
Glioma HINT         & 76\% & 2.60  & 0.25 & 0.09  & 6  \\ \hline
\rowcolor[HTML]{FFFFFF} 
Glioma IntAct       & 73\% & 3.35  & 0.22 & 0.07  & 6  \\ \hline
\rowcolor[HTML]{FFFFFF} 
Glioma Reactome     & 97\% & 1.92  & 0.67 & 0.27  & 4  \\ \hline
\rowcolor[HTML]{FFFFFF} 
Glioma STRING       & 100\% & 2.09  & 0.68 & 0.24 & 4  \\ \hline
\end{tabular}
\end{table}

The column LCC (\%) shows the percentage of nodes present in the largest connected component relative to genes in the KEGG's pathways. The cancer pathways for Colorectal, Pancreatic, and Glioma have respectively 86, 76, and 75 genes. It is a tiny set compared the to thousands of nodes presented in the PPINS. Although small in size, the cancer pathways are biologically significant and indicate important interactions within the cell. On average, the LCC contains 88\% of the genes associated with the selected cancer types in the KEGG's database. Reactome and STRING have the greatest overlap, while IntAct has the least. The column AvgSP is the average shortest path, AvgC is the average clustering, $\rho$ is network density, and Com is the number of communities. These four global measures show that the sub-networks from HINT and IntAct are similar. The same happens with the sub-networks from STRING and Reactome. The similarities follow the trend found in previous analyses, with the difference that cancer networks from STRING and Reactome are closer together than the actual PPINs. The AvgC is significantly higher in the cancer network than with PPINs, especially in HINT and IntAct, where global clustering is only 8\%. This behavior is expected, since pathways usually form modules with high clustering coefficient~\autocite{Network_analysis_2012_Tutorial}.

\begin{figure}[!htb]
    \centering
    \includegraphics[width=0.75\linewidth]{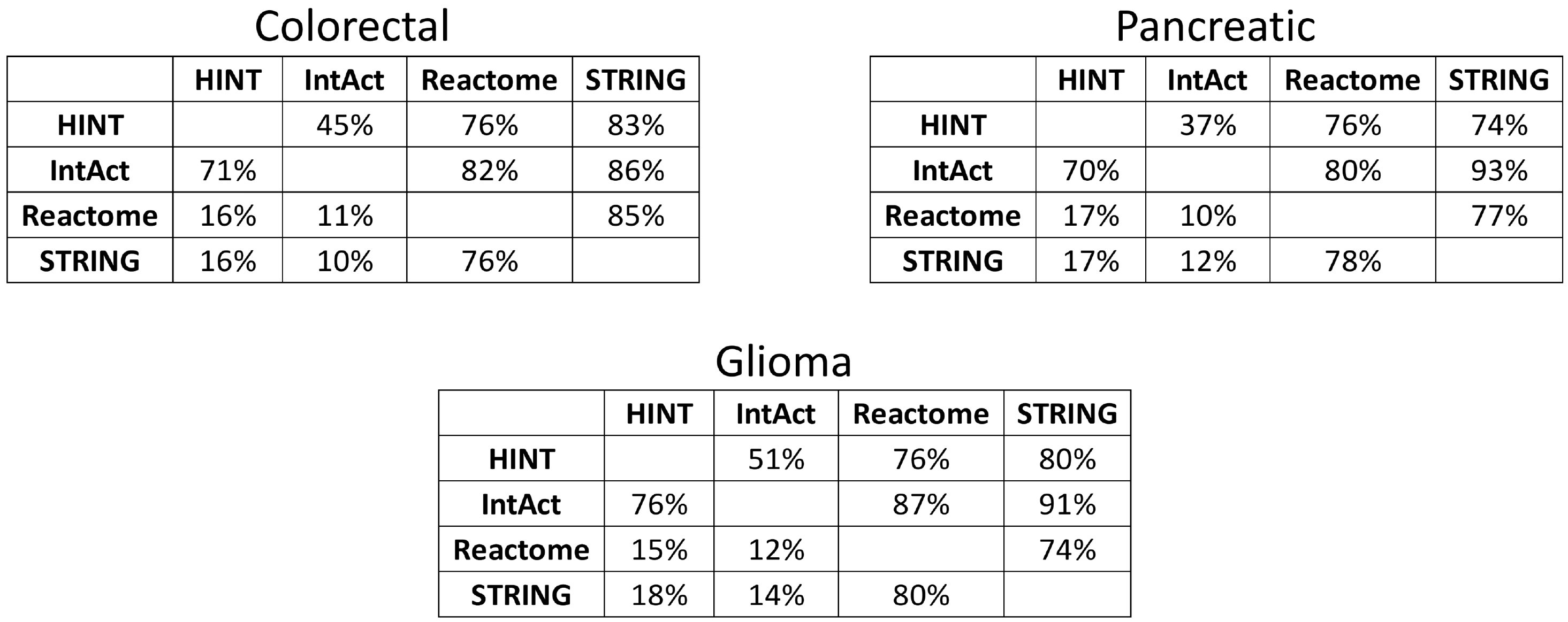}
    \caption{Cancer Pathways Networks: Edges Contained in Other Networks }
    \label{fig:edgeCancer}
\end{figure}

We replicate the analysis made in Table~\ref{tab:edgesIntersection} with the cancer networks. Figure~\ref{fig:edgeCancer} presents one table for each cancer type showing the percentual of edges from one network contained in the others networks. Compared to Table~\ref{tab:edgesIntersection}, we see a better overall intersection among the cancer network than with the whole networks. HINT decreases its association with IntAct from 62\% to an average of 44\% and significantly increases with Reactome and STRING, jumping from 8\% and 10\% to an average of 76\% and 79\%. IntAct also showed a similar increase with Reactome and STRING, but without decreasing it with HINT. Reactome and STRING also increase their mutual association from 40\% and 36\% to an average of 79\% and 78\%. 

The results shows that small sub-networks from cancer pathways enhance the similarity between PPINs using traditional measures and significantly increase edge intersections. Pathways are main functional modules within cells, and these results indicate that although the entire PPINs exhibit differences, mainly in edges, the similarity increases in pathway networks.

\section{Genes Centrality Role}
\label{sec:centralities}

\begin{figure}[!htb]
    \centering
    \includegraphics[width=0.95\linewidth]{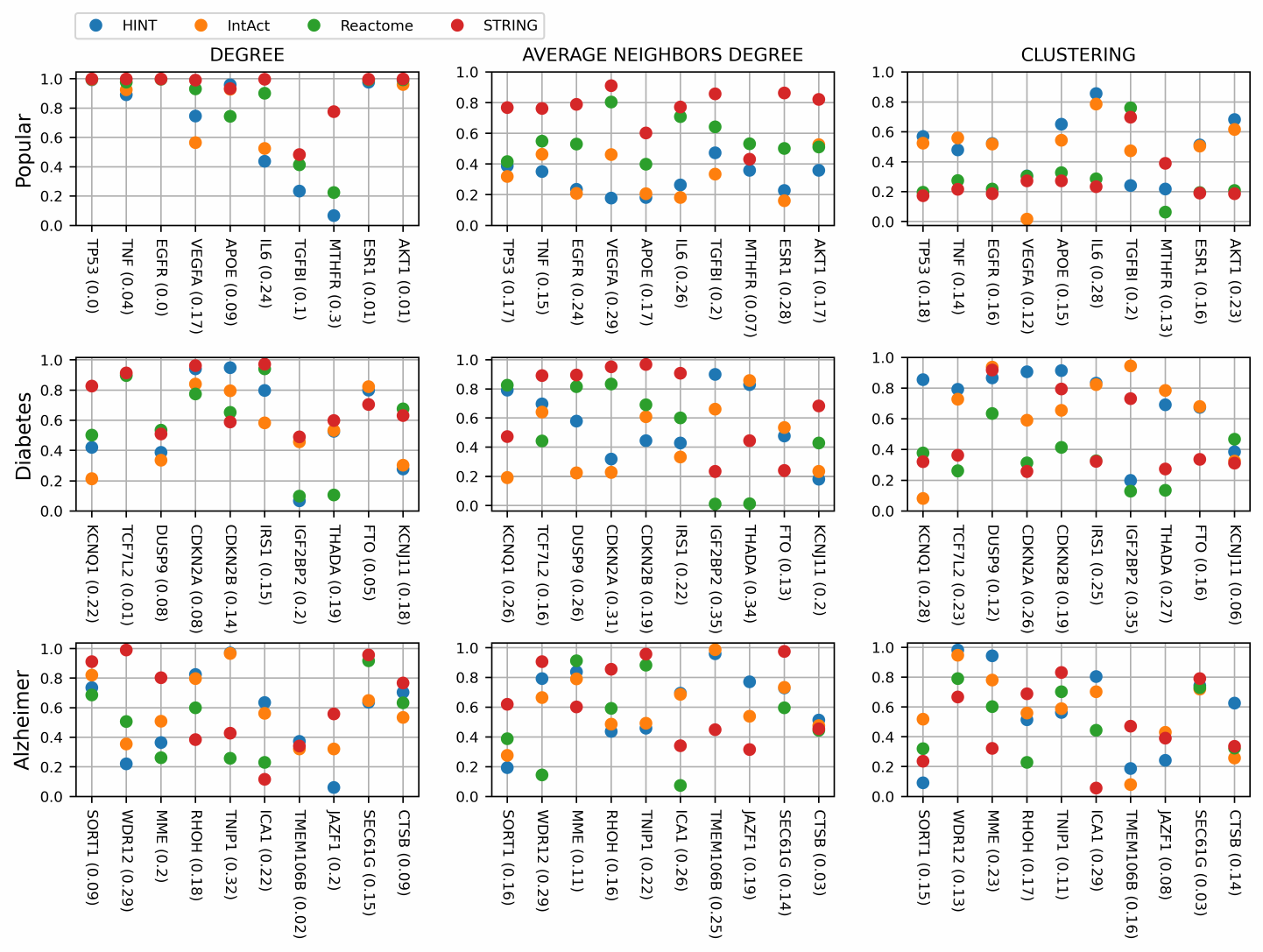}
    \caption{Percentile position for individual genes}
    \label{fig:targetGenes}
\end{figure}

In addition to conducting global and sub-network analyses, we also examine the topological role of three sets of nodes in the four PPINs. We have chosen 10 famous genes~\autocite{10popularGenes}, 10 genes associated with type 2 diabetes~\autocite{DiabetesGenes}, and 10 genes related to Alzheimer's disease~\autocite{AlzheimerGenes}. Figure~\ref{fig:targetGenes} shows their percentile positions in three centrality measures, the first indicating connectivity and the other two neighbourhood. The standard deviation is presented after the gene names, rounded to two decimal places.

In the first sub-plot, we present the percentile position in the degree distribution for the set of popular genes. The first gene, TP53, is a major hub in all PPINs. The percentile position for HINT, IntAct, Reactome and STRING, respectively, is 0.993, 0.994, 0.997, 0.999. Thus all points in the plot superpose. Something similar happens with the third gene, EGFR, and the last two genes, ESR1 and AKT1. The overlapping of points seen in these four genes indicates that, in terms of degree, these genes occupy the same topological role in the four PPINS. We would expect to find many overlappings in different measures and sets of genes if the PPINs were similar. However, Figure~\ref{fig:targetGenes} shows the opposite. Overall, the same gene has a spread percentile distribution among the PPINs. TP53, albeit a top hub in all PPINs, has a distinct neighbourhood in different PPINs. The percentile position for Average Neighbors Degree is 0.387 (HINT), 0.318 (IntAct), 0.415 (Reactome), and 0.767 (STRING). For clustering, the percentiles are 0.569 (HINT), 0.524 (IntAct), 0.195 (Reactome), and 0.173 (STRING). The genes TMEM106B, from Alzheimer, and  TCF7L2, from Diabetes, also have a similar degree and different neighbourhoods.

\section{Conclusion}

Considering that distinct databases make available PPINs for human interactome, this study aimed to topologically compare four networks: HINT, IntAct, Reactome, and STRING. Our analysis comprehends interaction significance, nodes and edges intersections, global measures, sub-networks comparisons, and node centrality. This coarse-to-fine approach offers a comprehensive overview, showing that the PPINs share some global characteristics, especially in cancer sub-networks, while differing in measures associated with neighbourhood and mainly in edges intersection.

Globally, the networks are scale-free, and the degree distribution in all of them follows the same trend. They are also small worlds and share similarities in diameter, average shortest path, eccentricity and community size. In the clustering distribution, HINT and IntAct are practically identical, with distant values from Reactome and STRING, which are similar to each other. The resilience to random and hub attacks followed the expected behaviour from scale-free networks, with STRING being the most resistant. HINT and IntAct perform similarly in assortativity, with a neutral behaviour in small degree nodes and a decreasing assortativity as the degree increases. Reactome showed a negative trend only with the highest hubs, with the other nodes being neutral. STRING differs from the others PPINs, with a positive assortativity in small degree nodes and mixed in hubs.

The significant differences in PPINs lay in the number of edges, edges weight, edges intersection and neighbourhood-associated measures. Even with the removal of low-score edges, the average of edges from one network contained in others is 22,75\%, and the average of nodes contained in others networks is 73,25\%. This shows that PPINs harbour many common protein-encoding genes but not their interactions. Centrality measures for the same genes also perform differently in the PPINs, even if the degree is similar. Sub-networks from cancer pathways showed to be more alike than the whole PPINs, showing the PPINs maintain topological similarities in functional modules.

The PPINs are evolving in an effort to model the human cell interactome, and albeit they share common nodes and some global characteristics, the interactions and associated measures differ. PPINs are used in many studies and computational analyses, like cancer driver genes discovery. Considering the results we present in this paper, we hypothesise that changing the PPIN may alter the result of previous findings. For future works, we propose to investigate this hypothesis and create a consensus PPIN that encompasses different PPINs.





\paragraph{\textbf{Data availability:}} All the code and data used in this work is available at: \url{https://github.com/RodrigoHenriqueRamos/Human-Protein-Protein-Interaction-Networks-A-Topological-Comparison-Review}.


\printbibliography




\end{document}